\documentclass[twocolumn,showpacs,preprintnumbers,amsmath,amssymb]
{revtex4}

\usepackage{graphicx}
\usepackage{dcolumn}
\usepackage{bm}

\begin{document}

\preprint{APS/123-QED}

\title{Mie-resonances, infrared emission and band gap of InN}

\author{T. V. Shubina}
\email{shubina@beam.ioffe.ru}
\author {S. V. Ivanov}
\author{V. N. Jmerik}
\author{D. D. Solnyshkov}
\author {V. A. Vekshin}
\author{P. S. Kop'ev}
\affiliation{Ioffe Physico-Technical Institute, Polytekhnicheskaya
26, St. Petersburg 194021, Russia}

\author{A. Vasson, J. Leymarie, and A. Kavokin}
\affiliation{LASMEA-UMR 6602 CNRS-UBP, 63177 AUBIERE CEDEX,
France}

\author{H. Amano} 
\author{K. Shimono}
\affiliation{Meijo University, 1-501 Shiogamaguchi, Tempaku-ku, Nagoya 468-8502, Japan}

\author{A. Kasic} 
\author{B. Monemar}
\affiliation{Dept. of Physics and Measurement Technology, Link\"oping University, S-581 83 Link\"oping, Sweden}

\date{\today}

\begin{abstract}
Mie resonances due to scattering/absorption of light in InN containing clusters of metallic In may have been erroneously interpreted as the infrared band gap absorption in tens of papers. Here we show by direct thermally detected optical absorption measurements that the true band gap of InN is markedly wider than currently accepted 0.7 eV. Micro-cathodoluminescence studies complemented by imaging of metallic In have shown that bright infrared emission at 0.7-0.8 eV arises from In aggregates, and is likely associated with surface states at the metal/InN interfaces. \end{abstract} 


\maketitle

InN is one of the most mysterious semiconductor materials studied up to now. While it has been carefully investigated since the 1980s using all the modern spectroscopic techniques, the band parameters of InN (gap energy, effective masses) are still under debate. This uncertainty is particularly embarrassing since InN-containing heterostructures have a huge potentiality for lasers, white-light diodes, and other devices. 

In this Letter we show that dramatic deviations in presently available experimental data on the band gap of InN from 0.7 to 2 eV \cite{1}-\cite{5} have a fundamental physical reason. They are linked to the precipitation of indium in the metallic phase that leads to additional optical losses associated with Mie resonances.

Resonant light scattering and absorption by dispersed small particles, described by Mie almost 100 years ago \cite{6}, have been thoroughly investigated in different objects in nature - cosmic dust, comets, atmosphere clouds $\it{etc.}$ (see, e.g., Ref. \cite{7}). They have also been repeatedly observed in various dielectrics and semiconductors with clusters of different metals \cite{8},\cite{9}. These phenomena arise from interaction of an incident electromagnetic wave with multipolar excitations of electrons in the clusters. Radiative processes associated with metallic inclusions have been continuously reported as well \cite{10}.

\begin{figure}
\includegraphics{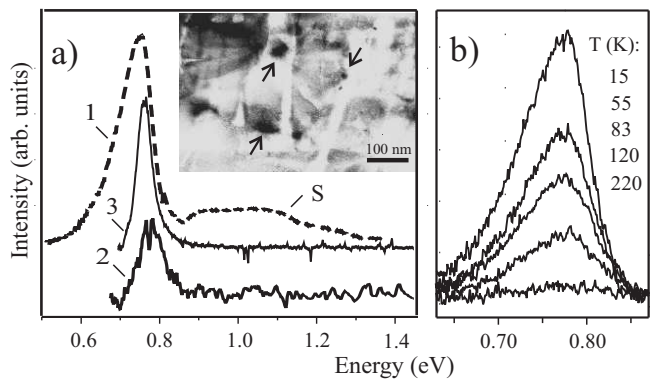}
\caption{\label{fig:epsart}
(a) Spectra of (1) PL excited by an InGaAs laser in an InN sample grown by MBE (registration by a combination of InGaAs and PbS detectors); $\mu$-CL spectra registered in (2) the same MBE sample and (3) a sample grown by MOCVD. The Ge detector cuts the long-wavelength tail of the CL spectra. The inset presents a cross-sectional TEM micrograph of an MBE sample with In precipitates marked by arrows. "S" indicates scattered signal (see text). (b) PL temperature behavior in the MBE sample (PbS detector). }
\end{figure}

Such effects have been neglected in studies of InN, although it is well known that the poor thermal stability of InN and low In vapor pressure over the metal phase result in formation of In clusters at temperatures higher than 500$^\circ$C, commonly used to obtain high quality InN \cite{5}. A good correlation between the recently registered infrared (IR) photoluminescence (PL) and the optical absorption edge has been considered as a doubtless evidence of the narrow gap \cite{3}-\cite{5}. An attempt has been made  to verify the failure of the band gap-common-cation rule \cite{11}, accordingly to which $E_g$ of InN has to be wider than that of InP (1.42 eV). 

\begin{figure}
\includegraphics{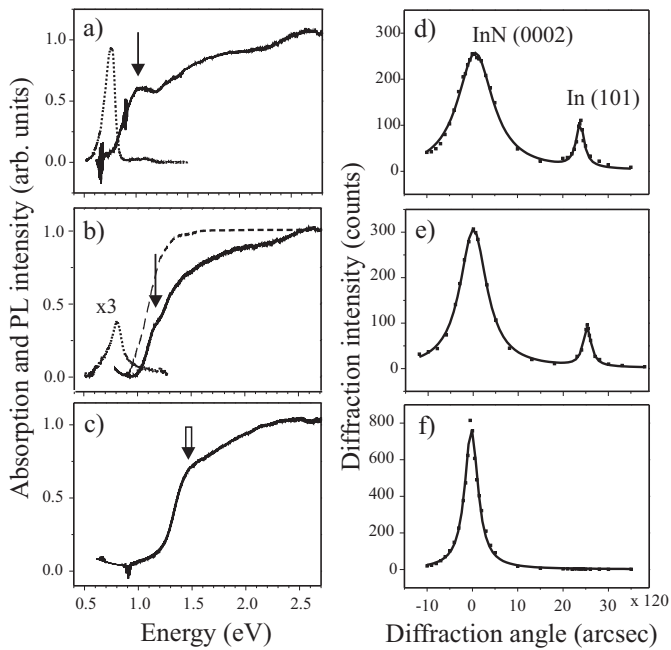}
\caption{\label{fig:epsart} 
TDOA (solid lines), IR PL (dotted lines), and optical absorption (dashed line) spectra taken in three representative MBE grown InN films with (a) high, (b) moderate and (c) negligible concentration of metallic In as registered by XRD. Respective XRD scans are shown in (d), (e), and (f). Vertical single arrows indicate an additional absorption peak (see the text); the double arrow denotes the position of the InN band edge, as determined by the TDOA technique. }
\end{figure}

We have examined two representative sets of InN epilayers grown by both plasma-assisted molecular beam epitaxy (MBE) and metalorganic chemical vapor deposition (MOCVD) on sapphire and other substrates. The maximum of the IR emission in these samples has been registered in the 0.7-0.8 eV range independently of the growth technique and the way of excitation, i.e., optically, by different laser lines, or by an electron beam in the cathodoluminescence (CL) study (Fig. 1). The latter has been performed in a LEO 1550 Gemini analytical scanning electron microscope (SEM) equipped with a liquid-helium high-resolution CL system. 

The electron concentration in the samples, as determined by IR ellipsometry measurements, varies from 2.1 up to 8.4$\times$10$^{19}$ cm$^{-3}$, assuming an effective electron mass of $m_{e}^{\ast}$ = 0.11$m_{0}$ \cite{12}. No noticeable correlation between the IR emission energy and the carrier concentration was observed. A shift of the PL with a temperature rise is very small (Fig. 1 (b)), similar to previous reports \cite{13}. 

The samples under study may be divided into three groups with (A) bright, (B) weak and (C) negligible IR emission (Fig. 2 (a-c)). Among ten MBE layers only one, grown with exceeding of 500$^\circ$C, belongs to group A. The most of the samples either does not emit light at all or show a very weak signal, in spite of reasonable structural quality. A similar picture has been observed in the MOCVD set, where two samples of eight, possessing bright IR emission, were grown directly on sapphire in the 500-550$^\circ$C temperature range. 

All the samples exhibiting the IR PL contain inclusions of a tetragonal In phase in hexagonal InN, as revealed by x-ray diffraction (XRD) measurements, performed using a sensitive triple-crystal diffractometer. As can be seen in Fig. 2, the PL correlates with the (101) In peak; both are most intense in sample A and disappears completely in the MBE samples C. Note that the In precipitates have been directly observed in a transmission electron microscopy (TEM) image as well (Fig. 1, the inset). 

We have studied whether there is any spatial correlation between the IR emission and the distribution of metallic In. This has been done in the same analytical microscope using a detector of backscattered electrons (BSE), sensitive to atomic weight, and an energy dispersive x-ray (EDX) analyzer. Figure 3 (a-d) summarizes results of the studies for the MBE sample A. It turns out that the IR CL arises within the In-enriched regions, visible as bright areas in both BSE image and mapping of In x-ray fluorescence. The IR emission is dotty, while emission from sapphire (not presented here), penetrating between the In-enriched regions at $\sim$8 kV, is homogeneous. Moreover, the spots of the IR emission coincide most frequently with specific defects, well resolved in the secondary electron (SE2) images. Such defects are, in fact, dips in the InN films above the contacts with the metal particles. The locally increased growth temperature induces dissociation of InN there and, hence, a reduction of the growth rate. 

In samples B, with the smaller In content, the distribution of IR emitting spots in the CL images is more homogeneous. Their density and average size ($\sim$100 nm) are smaller than in the sample A. These spots appear on the In agglomerations which have almost circular shapes in the BSE images. In the samples C, we failed to register the IR emission at all. The BSE images of such layers do not exhibit any noticeable amount of the metallic In either, and accordingly the dips are absent in the respective SE2 images (Fig. 3 (e, f)). 

In the MOCVD samples, the metallic In tends to agglomerate at the boundaries of hexagonal grains, as can be seen in both SE2 and BSE images (Fig. 4 (a, b)). Accordingly, the IR CL looks like a fine net along the grain boundaries (Fig. 4 (c, d)). With an increase of the electron beam energy, the net becomes brighter, but no noticeable emission appears inside the hexagons. The MOCVD samples demonstrate a similar disappearance of the IR PL in films without metallic In. Obviously, the bright IR (0.7-0.8 eV) emission in both sets of the InN samples is associated with the In aggregates.

\begin{figure}
\includegraphics{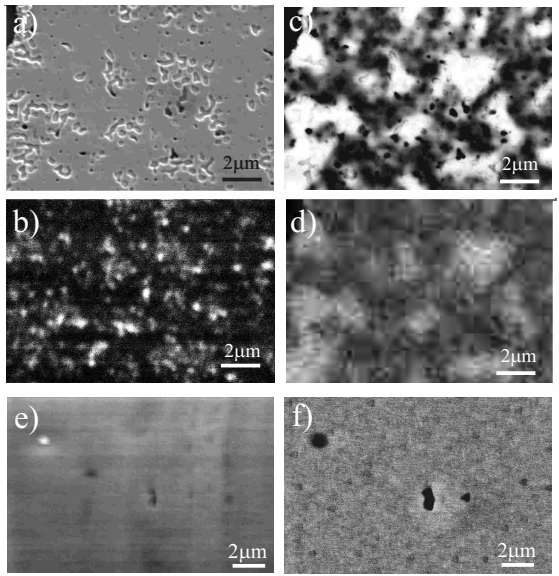}
\caption{\label{fig:epsart} Images registered at the same spot of the MBE sample A: (a) SE2 at 5 keV; (b) CL at 4.3 kV, (c) BSE at 20 keV, (d) In mapping in an energy window of 3.287-3.487 kV (InLa1 and InLb1). White regions in (c) and (d) are In enriched. Images (e) SE2 and (f) BSE are taken at 20 keV in the MBE sample C where the IR CL has not been registered. 
}
\end{figure}

\begin{figure}
\includegraphics{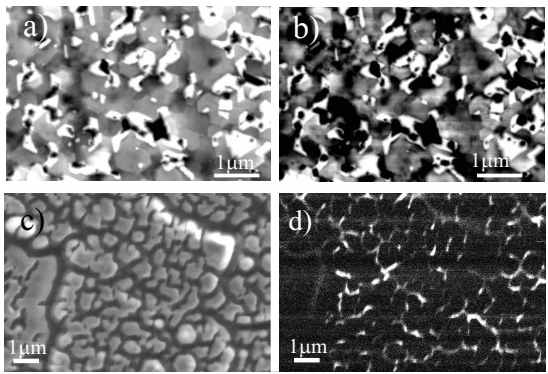}
\caption{\label{fig:epsart} 
Typical (a) SE2 and (b) BSE images registered at 20 keV in the MOCVD sample A demonstrating the IR PL. White inclusions between hexagons are metallic In. (c) SE2 and (d) CL images taken at 4.85 kV in this area. 
}
\end{figure}

Total optical extinction losses in a semiconductor matrix with metallic clusters, besides the interband absorption in the matrix, contain two additional components: i) true (bipolar) absorption or transformation of radiation energy into heat in small particles, and ii) resonant scattering on plasmon excitations, which is important for larger particles \cite{8}. Characteristic features of both components are observed in our optical spectra. 

The PL spectra often contain an additional "S" band above the main one, in the 0.8-1.4 eV range (Fig. 1 (a)). Our study has shown that this is the contribution of an abnormally strong scattered background signal (non-monochromatic spontaneous emission of the semiconductor laser and/or fluorescence of all optical components at high excitation power). Apparently, the scattered signal is absent in the CL spectra due to the electron beam excitation without such a background. At the normal incidence, the signal may be so strong that it masks the real IR emission. This scattering can falsify the results of conventional optical absorption measurements as well, since less optical energy reaches the detector.

To distinguish between such optical losses and true absorption we have used the thermally detected optical absorption (TDOA) technique. This method, performed at 0.35 K, is based on the detection of a small increase in the sample temperature, which is proportional to true absorption \cite{14}. In our case, the increase arises from the creation of phonons at nonradiative recombination activated by the initial interband optical absorption and from the bipolar absorption of light in the metallic particles. 

The difference between the TDOA and the conventional absorption is illustrated by spectra in Fig. 2 (b). The optical absorption extends to lower energies and completely saturates at higher energies. In contrast, the TDOA features a complicated spectrum with a pronounced additional peak below a principal absorption edge. This peak shifts to the higher energy in the spectra of the samples B and disappears totally in the samples C, where In clusters are not detected. Therefore, we consider the peak to be associated with the absorption within the In particles. 

In the classical Mie theory, the extinction losses of a metallic sphere embedded in a matrix are expressed through an imaginary part of its polarizability. This quantity depends on the complex dielectric functions of both matrix material ($\epsilon$) and metal ($\epsilon_m$). As shown by qualitative EDX analysis, there is excess of In of a few percents in our samples. Such a filling factor $f$ should significantly change the complex dielectric function of any semiconductor. Besides, the shape of In inclusions is far from spherical. To estimate the resonant absorption energies, we have used the model, based on the Maxwell-Garnett approximation for an effective medium, developed for non-spherical metal particles \cite{15},\cite{9}. In this model, the absorption is given as 

\begin{equation}
\alpha\propto{\rm Im}[\epsilon(w)G(w)],
\nonumber
\end{equation}  
where

\begin{equation}
G(w)=(1+f\frac{(\epsilon_{m}-\epsilon)(1-L_m)}
{L_m\epsilon_{m}+(1-L_m)\epsilon})/
(1-f\frac{(\epsilon_{m}-\epsilon)L_m}
{L_m\epsilon_{m}+(1-L_m)\epsilon}).
\nonumber
\end{equation}  

Here $L_m$ denotes the depolarization factor, which equals $1/3$ for spheres. The dielectric function for In is approximated by the Drude model. We take the material parameters of In from Refs. \cite{16}, \cite{17}, and the complex dielectric function of InN from Ref. \cite{18}.
 
Figure 5 demonstrates that this model enables one to describe the TDOA data on absorption in InN. Reasonable agreement between the experimental peak and the calculated Mie resonance for the MBE sample A is obtained with $f=2\%$ and $L_m=0.066$ (Fig. 5 (a)). The blue shift with the decrease of the In content from 4 to 0.5$\%$ is 0.4 eV. Deviation from the spherical shape provides an even more crucial effect, as can be seen in Fig. 5 (b), where only the normalized Mie resonances are presented for the sake of demonstrativeness (interband transitions cause the damping of the resonances above 2 eV).

\begin{figure}
\includegraphics{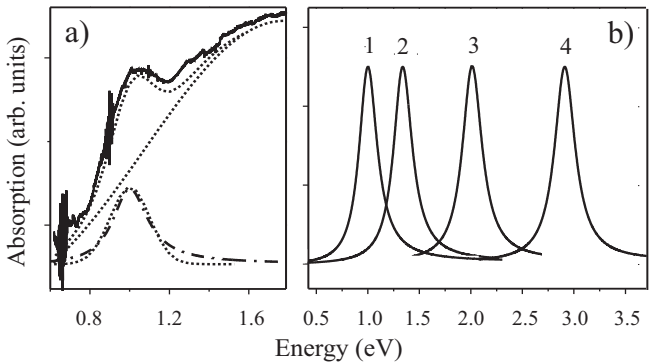}
\caption{\label{fig:epsart} 
(a) The experimental TDOA spectrum of the MBE sample A (solid line) with its Gaussian deconvolution (dotted lines) to show an additional absorption peak. Fitting of the peak (dash-dotted line) by the Mie resonance is performed with $f=2\%, L_m=0.066$. (b) Calculated absorption in InN containing $2\%$ of In at (1) $L_m=0.066$, (2) $L_m=0.1$, (3) $L_m=0.2$, and (4) $L_m=0.33$.}
\end{figure}

One should clarify that the band gap energy in a TDOA spectrum corresponds to the position of the kink between the constant and slope parts of the spectrum. The kink position in the spectra of InN samples, having no detectable metallic inclusions, is near 1.4-1.5 eV. Therefore, it is tempting to conclude that the true band edge of InN is in this range, with possible correction on the Burstein shift \cite{19}. However, it is difficult to separate the interband absorption in InN from the Mie absorption in ultra small In nanoclusters. Thorough high-resolution TEM studies are needed to determine finally this fundamental characteristic. 

The IR emission is situated significantly below the principal edge in TDOA spectra and below the Mie resonances. It arises on an interface of InN with a large enough In particle ($>$100 nm). Among the possible mechanisms, recombination involving surface states at an interface metal/semiconductor seems to us the most reasonable. Optical emission in several III-V compounds with thin metal coverage had been observed $\sim$0.7 eV below the respective band edge \cite{20}. In perfect crystals, like the InN hexagons, the Tamm surface states can appear due to the change in the periodic crystal potential at an interface \cite{21}. The potential change is essential, if a metal inclusion is "bulk-like", i.e., with a size much larger than the exciton Bohr radius. The energy of such states, governed by the Schottky-barrier height and the Fermi level stabilized by the metal inclusions, is weakly sensitive to the band energy variation with temperature, in agreement with our observation. 

In conclusion, the presence of In precipitates could explain a lot of experimental data: the huge band edge shift, variation of the sample coloration from red to metallic, and the IR emission and scattering. The Mie resonances are important for InGaN epilayers as well, because of the observed activation of In segregation phenomena in the alloys at high growth temperatures. We resume that InN of high optical quality, grown at $T>500^\circ$C, absorbs and emits light far below its bandgap due to Mie resonances and Tamm states, respectively. 

We thank for help in structural characterization Dr. V. V. Ratnikov, V. M. Busov, and A.A. Sitnikova. This work was partly supported by RFBR (grants 03-02-17563 and 03-02-17567) and the cooperative Grant between Russia and France N04509PB. A. Kasic acknowledges support from The Wenner-Gren Foundation, Sweden.



\end{document}